\documentclass[twocolumn,showpacs,amsmath,amssymb,pra]{revtex4}

\usepackage{graphicx}
\usepackage{color}
\usepackage{dcolumn}
\usepackage{amsmath}
\usepackage{CJK}
\usepackage{subfigure}

\usepackage{exscale}
\usepackage{relsize}
\usepackage{bm}

\usepackage{mathrsfs}
\usepackage{amsfonts}
\usepackage{indentfirst}

\newcommand{\be}{\begin{equation}}
\newcommand{\ee}{\end{equation}}
\newcommand{\bey}{\begin{eqnarray}}
\newcommand{\eey}{\end{eqnarray}}
\newcommand{\bw}{\begin{widetext}}
\newcommand{\ew}{\end{widetext}}

\newcommand{\ra}{\rangle}
\newcommand{\la}{\langle}

\newcommand{\ba}{\begin{array}}
\newcommand{\ea}{\end{array}}
\newcommand{\bi}{\begin{itemize}}
\newcommand{\ei}{\end{itemize}}
\newcommand{\bem}{\begin{enumerate}}
\newcommand{\eem}{\end{enumerate}}

\begin{document}

\title{Understanding quantum work in a quantum many-body system}

\author{Qian Wang$^{1}$ and H.~T.~Quan$^{1,2}$\footnote{ Electronic address: htquan@pku.edu.cn}}

\affiliation{$^{1}$School of Physics, Peking University, Beijing 100871, China \\
 $^2$Collaborative Innovation Center of Quantum Matter, Beijing 100871, China}

\date{\today}

\begin{abstract}
  Based on previous studies in a single-particle system in both the integrable [Jarzynski, Quan, and Rahav, Phys.~Rev.~X {\bf 5}, 031038 (2015)]
  and the chaotic systems [Zhu, Gong, Wu, and Quan, Phys.~Rev.~E {\bf 93}, 062108 (2016)],
  we study the the correspondence principle between quantum and classical work distributions in a quantum many-body system.
  Even though the interaction and the indistinguishability of identical particles increase the complexity of the system,
  we find that for a quantum many-body system the quantum work
  distribution still converges to its classical counterpart in the semiclassical limit.
  Our results imply that there exists a correspondence principle between quantum and classical
  work distributions in an interacting quantum many-body system, especially in the large particle number limit,
  and further justify the definition of quantum work
  via two-point energy measurements in quantum many-body systems.

\end{abstract}


\maketitle

\section{Introduction}
 In recent years, the field of nonequilibrium statistical mechanics in small systems \cite{cbu,seifert,jarzynski,rkw} has attracted lots
 of attention. A major breakthrough in this field in the past two decades is the discovery of exact fluctuation
 relations, which hold true for systems driven arbitrarily far from equilibrium.
 Their validity has been confirmed in various experimental and numerical studies \cite{jarzynski1,jarzynski2,crooks1,crooks2,crooks3,usf,meu,kck,gzp}.
 Now, these relations are collectively known as fluctuation theorems (FTs).
 The FTs have provided insights into the physics of nonequilibrium
 processes in small systems where fluctuations are important \cite{jarzynski}.
 Despite these great developments,
 there are still some aspects of these FTs that have not been fully understood.
 The definition of the quantum work is one example. There have been many definitions
 of quantum work for an isolated system \cite{ppt}.
 However, only the work defined through two projective measurements of the system's instantaneous energy, i.e.,
 at the start ($t=0$) and at the end ($t=\tau$) of the driving process \cite{meu,pte,mcp,bpv,bgp,kur,tak}, satisfies the FTs.
 Although this definition of quantum work satisfies quantum nonequilibrium work relations, it might seem {\emph{ad hoc}}.
 This is because the collapse of the wave function \cite{vonn}, when measuring the final energy, brings profound interpretational
 difficulty to the definition of quantum work \cite{chs}.
 Therefore, it is necessary to find other independent evidence (besides the validity of the FTs) to
 justify the definition of quantum work via two-point energy measurements.

 Recently, the quantum work defined via two-point energy measurements has been justified
 in both a one-dimensional integrable system \cite{chs} and a chaotic system \cite{zhulong,mata} through
 the correspondence between quantum and classical work distributions.
 By using the semiclassical method \cite{littlejohn,delos} and the numerical simulation, it is shown that in the
 semiclassical limit, i.e., $\hbar\to0$,
 the quantum work distribution converges to the classical work distribution after ignoring the effect due to interference of classical trajectories \cite{chs}.
 Therefore, there is a quantum-classical correspondence principle of work distributions.
 Thus, these studies provide some justification to the definition of the quantum work,
 because the classical work is well defined without any ambiguity.
 Nevertheless, for quantum many-body systems, the correspondence between quantum and classical work distributions has
 not been studied so far.
 The indistinguishability of identical particles \cite{gong,tichy} and the interaction makes the properties of quantum work even more elusive.
 Also, the nonequilibrium dynamic evolution of a quantum many-body system is extremely difficult to solve.
 Following a similar argument to that in Refs.~\cite{kur,tak}, it can be checked that quantum work
 defined via two-point energy measurements in a quantum many-body system satisfies FT.
 For example, the work fluctuations in bosonic Josephson junctions has been studied in Ref.~\cite{rggm}.
 But a deeper understanding about quantum work in a quantum many-body system is still lacking.
 And the quantum work mentioned above has not been justified in these systems.
 In this article we aim to explore the properties of quantum work in a quantum many-body system, i.e., a one-dimensional (1D) Bose-Hubbard (BH) model,
 and study the correspondence principle of work distributions when both indistinguishability and interaction play an important role.

 The BH model which describes an interacting boson in a lattice potential£¬
 constitutes one of the most extensively studied and most fundamental
 Hamiltonians in the field of condensed matter theory and quantum simulation.
 It undergoes a transition from a superfluid phase to an insulator phase as the strength of the potential is increased
 \cite{fisher,sachdev,kfh,neh,avp,dcb,apk,cok,mgo}.
 Meanwhile, this quantum many-body system has a classical limit.
 The classical limit of this model is described by the celebrated discrete nonlinear Sch\"{o}dinger equation \cite{jce},
 which possesses rich properties in both static and dynamic aspects.
 We study the work distribution of this system in both quantum and classical regimes.
 The results show that there indeed exists a quantum-classical correspondence between work distributions in this quantum many-body system.
 Furthermore, we investigate when the correspondence principle between work distributions
 will break down with the decrease of the number of particles.
 Our study justifies the definition of the quantum work via two-point energy measurements in a quantum many-body system.

 The remainder of this article is organized as follows. In Sec.~\ref{BHM}, we introduce the 1D BH model, briefly review its properties,
 and discuss the classical limit of it.
 The quantum and classical work distributions are compared in
 Sec.~\ref{QCTP} where we prove that the correspondence principle between quantum and classical work distributions can be reduced to
 the correspondence principle between the quantum and classical transition probabilities.
 Then we give definitions and discussions of the quantum and classical transition probabilities between different energy eigenstates.
 Our numerical results and analysis are provided in Sec.~\ref{NRS} where we show that the quantum and classical transition
 probabilities in the 1D two-site and three-site BH models converge in the semiclassical limit.
 Finally, conclusions and discussions are given in Sec.~\ref{DSC}.

 \begin{figure}
  \includegraphics[width=\columnwidth]{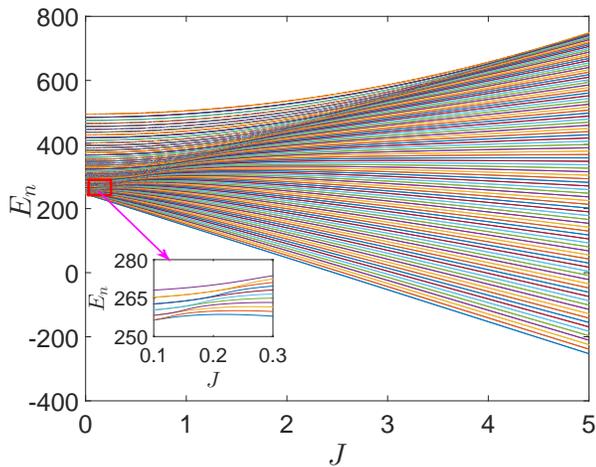}
  \caption{(Color online) Energy spectrum of the 1D two-site BH Hamiltonian (\ref{twoBH}) as a function of the work parameter $J$ for $N=100$.
   Inset: The details of the energy spectrum in the red rectangle.}
  \label{energy}
 \end{figure}

\section{1D Bose-Hubbard model} \label{BHM}

 The Hamiltonian of the standard 1D BH model is written as
 \be \label{BHH}
    \hat{H}=\sum_j^L\left[-J(\hat{a}_j^\dag\hat{a}_{j+1}+\hat{a}_{j+1}^\dag\hat{a}_{j})+
         \frac{U}{2}\hat{a}^\dag_j\hat{a}_j(\hat{a}^\dag_j\hat{a}_j-1)\right],
 \ee
 where $\hat{a}_j, \hat{a}_j^\dag$ are bosonic annihilation and creation operators for the $j$th site and
 satisfy the usual bosonic commutation rules $[\hat{a}_i, \hat{a}_j^\dag]=\delta_{ij}$ and $L$ denotes the number of sites.
 $U$ is a measure for the on-site two-body interaction strength depending on the $s$-wave scattering length,
 and $J$ denotes the tunneling amplitude, which depends on the barrier height \cite{mgo,dpz}.
 Here the periodic boundary condition, i.e., $a_{L+1}=a_1$, has been assumed.
 Obviously, it is straightforward to check that $[\hat{H}, \hat{N}]=0$ with $\hat{N}=\sum_j\hat{a}_j^\dag\hat{a}_j$.
 The total number of particles $N=\sum_j n_j$ is a conserved quantity, and the dimension of the Hilbert space is
 $\mathrm{dim}[H]=C_{N+L-1}^N$.
 Such a model can be experimentally realized by using cold atoms in an optical lattice \cite{cok,mgo,amico,tpm,dpz}.

 The interactions between the bosons can be characterized by a dimensionless coupling parameter \cite{apk,mck,lena,lena1,hvs,gsp,leggett}
 \be \label{Lambda}
     \lambda=\frac{UN}{J}.
 \ee
 For the two-site case, depending on the values of $\lambda$, one can identify three qualitatively different regimes \cite{hvs,leggett,gsp,mck,lena,lena1}.
 The Rabi regime ($\lambda<1$), the Josephson regime ($1<\lambda\ll N^2$), and the
 Fock regime ($\lambda\gg N^2$).
 Due to the interplay between the tunneling and the on-site interaction among the bosons, the
 BH model exhibits rich and interesting dynamical properties.

 \begin{figure*}
  \centering
  \subfigure[]{\label{fig:subfig:a}
  \includegraphics[width=0.48\textwidth]{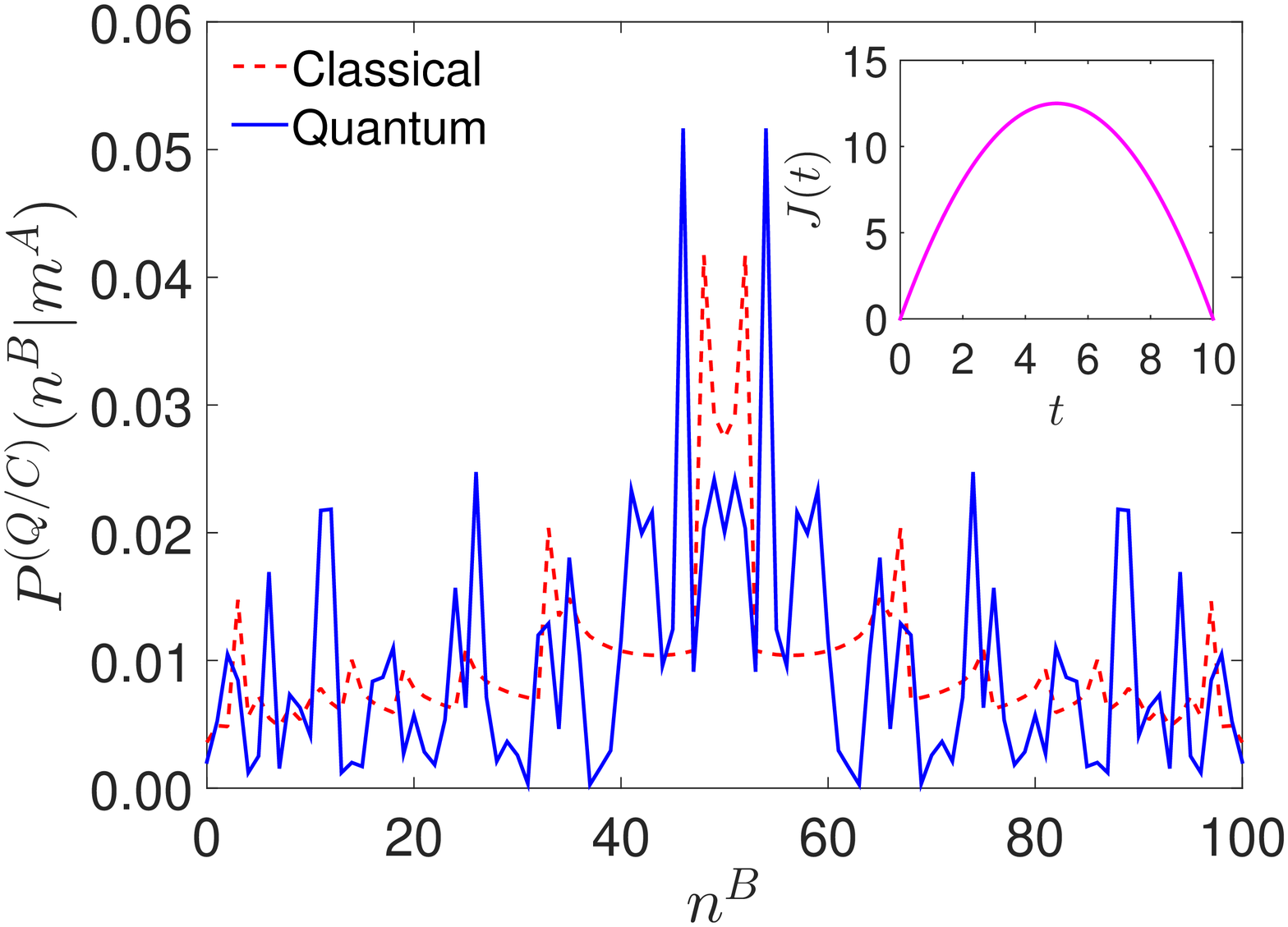}}
  \subfigure[]{\label{fig:subfig:b}
  \includegraphics[width=0.48\textwidth]{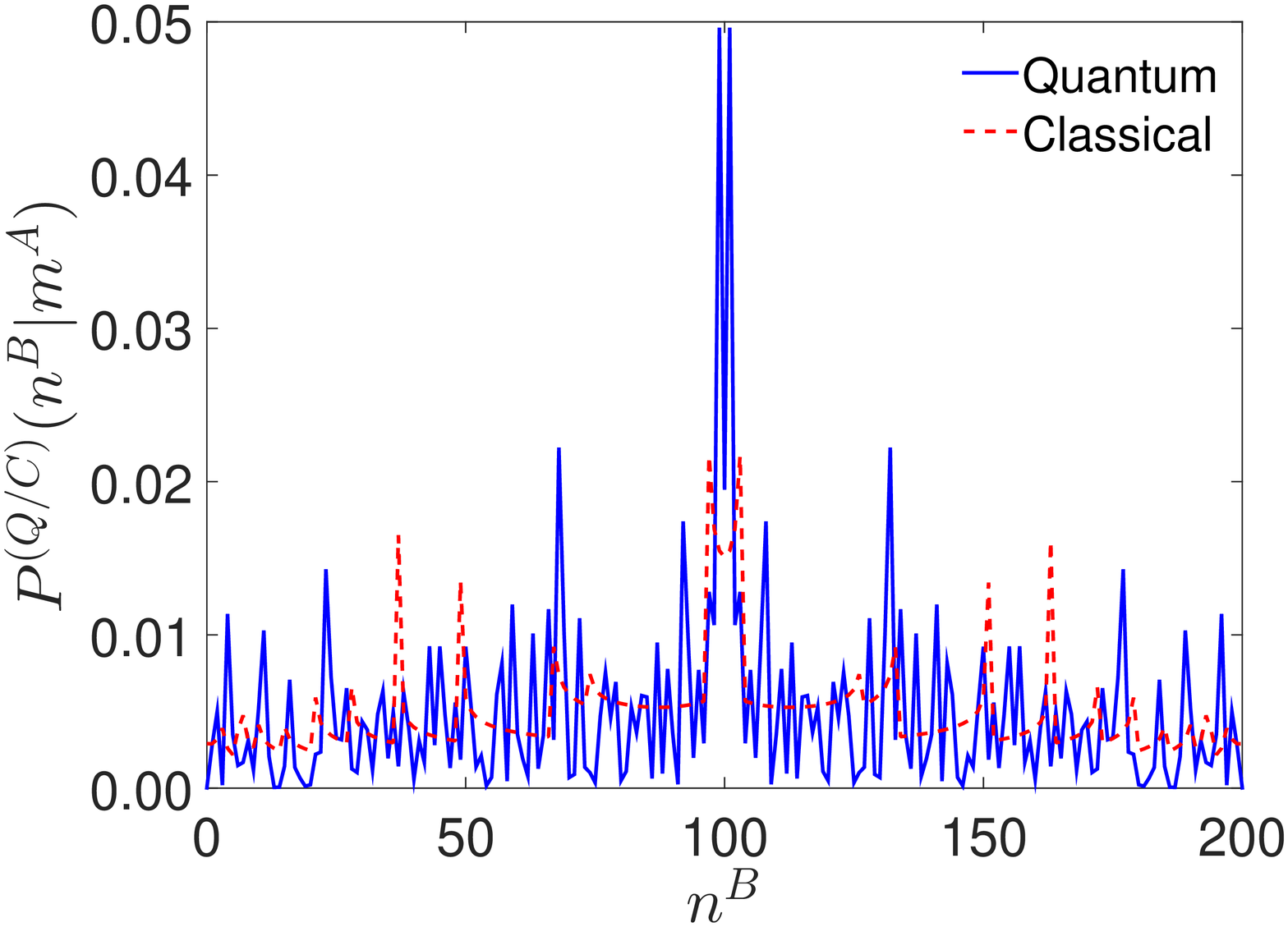}}
  \caption{(Color online) Quantum [Eq.~(\ref{QTPF})] and classical [Eq.~(\ref{RCp})] transition probabilities for
  the 1D two-site BH model with different number of particles (a) $N=100$, (b) $N=200$. The solid blue curve represents the quantum case $P^Q(n^B|m^A)$,
  while the dashed red curve represents the classical case $P^C(n^B|m^A)$. For the quantum case, the initial state is the ground state
  of $H(t=0)$ with $|m^A\ra=|\mathbf{n}^A\ra=|N/2,N/2\ra$.
  For the classical case, the initial state is a collection of microscopic states $\Psi^A=\{\psi_1^A,\psi_2^A\}=\{(N/2,\phi_1^A),(N/2,\phi_2^A)\}$,
  with $\phi_1^A$ and $\phi_2^A$ the independent random numbers
  evenly sampled in the range $[0,2\pi)$.
  Here, due to $J(t=\tau)=0$, the finial energy eigenstates are given by the Fock states: $|n^B\ra=|n_1^B,N-n_1^B\ra$ with $n_1^B=0,1,\ldots,N$.
  Inset: Time dependence of the work parameter $J(t)$.}
  \label{qcp}
 \end{figure*}

 The semiclassical limit of this model can be achieved when $N\to\infty$; in other words, the effective Planck constant is given by $\hbar_{\mathrm{eff}}=1/N$ \cite{englt}.
 With $N\to\infty$, one can
 replace the annihilation and creation operators by complex numbers
 \cite{leggett,apk,heis,emg,kwm,aap,ass,sra,rfv,smc,ark,engl,englt}:
 \be \label{trf}
   \hat{a}_j\to\psi_j,\quad \hat{a}^\dag_j\to\psi_j^\ast,
 \ee
 with
 \be \label{clFS}
    \psi_j=\sqrt{n_j+\frac{1}{2}}\exp\{i\phi_j\}.
 \ee

 Then, one finds that the classical counterpart of the Hamiltonian (\ref{BHH}) is given by
 \be \label{MFH}
   \mathcal{H}_c=\sum_j^L\left[-J(\psi_j^\ast\psi_{j+1}+\psi_{j+1}^\ast\psi_j)+\frac{U}{2}|\psi_j|^4\right],
 \ee
 with Poisson brackets
 \be
    \{\psi_i,\psi_j^\ast\}=\delta_{ij},
 \ee
 and
 \be
     \{\mathcal{H}_c,\mathcal{N}\}=0,
 \ee
 where $\mathcal{N}=\sum_j|\psi_j|^2=N+L/2$ \cite{emg,smc,engl,englt}.
 The time evolution of the complex valued mean-field amplitudes $\psi_j$ are given by the following
 equation \cite{aap}:
 \bey
    i\hbar\frac{\partial\psi_j}{\partial t}=\frac{\partial\mathcal{H}_c}{\partial\psi_j^\ast}=-J(\psi_{j+1}+\psi_{j-1})+U|\psi_j|^2\psi_j. \label{CTE}
 \eey
 This equation can be regarded as the Hamilton equation of the mean field $\psi_j$.

 In the following sections, we will study the quantum-classical correspondence of work distributions in the 1D BH model
 based on the quantum and classical pictures given above.

 \begin{figure*}
  \centering
  \subfigure[]{\label{fig:subfig:a}
  \includegraphics[width=0.485\textwidth]{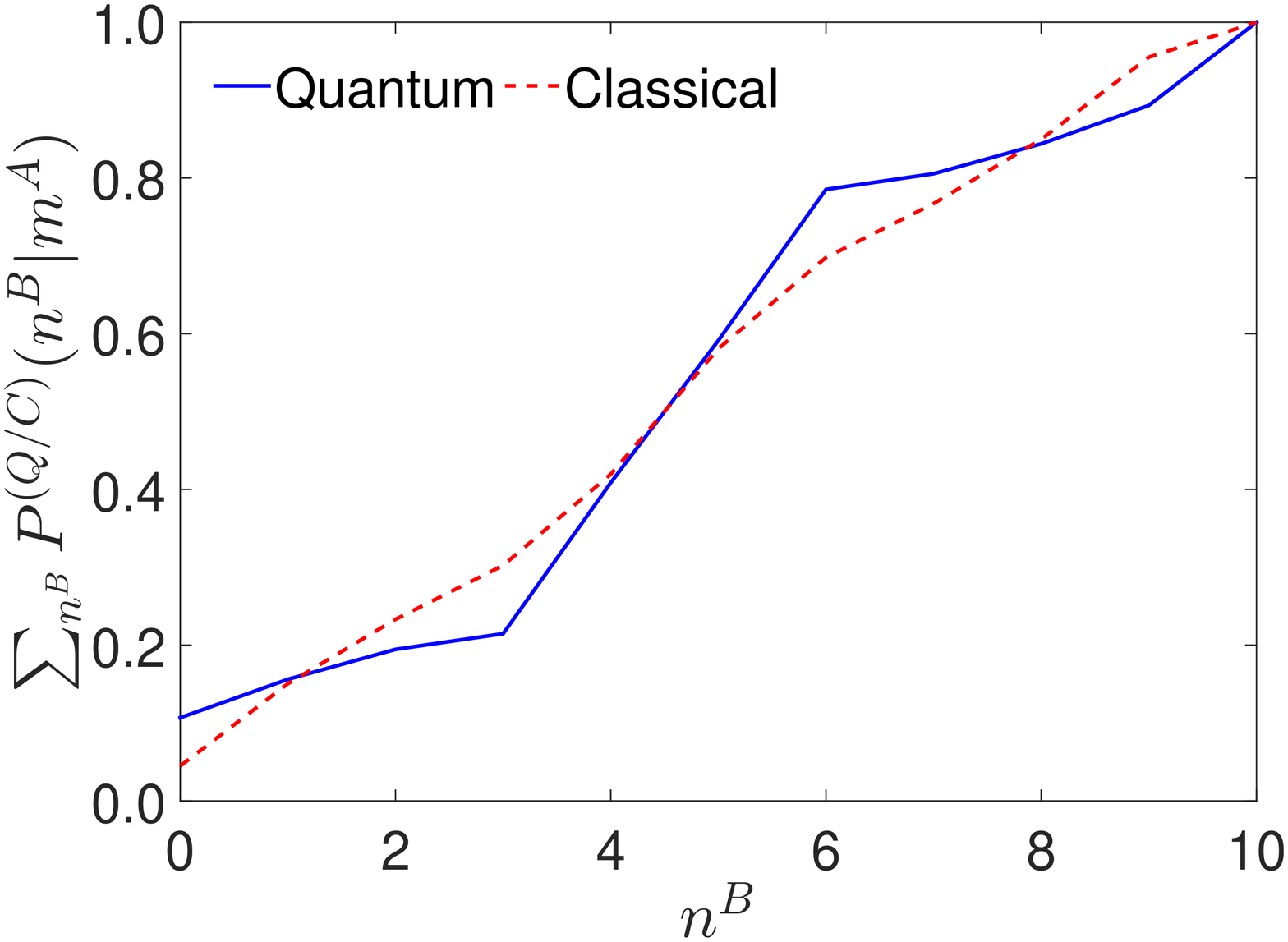}}
  \subfigure[]{\label{fig:subfig:b}
  \includegraphics[width=0.485\textwidth]{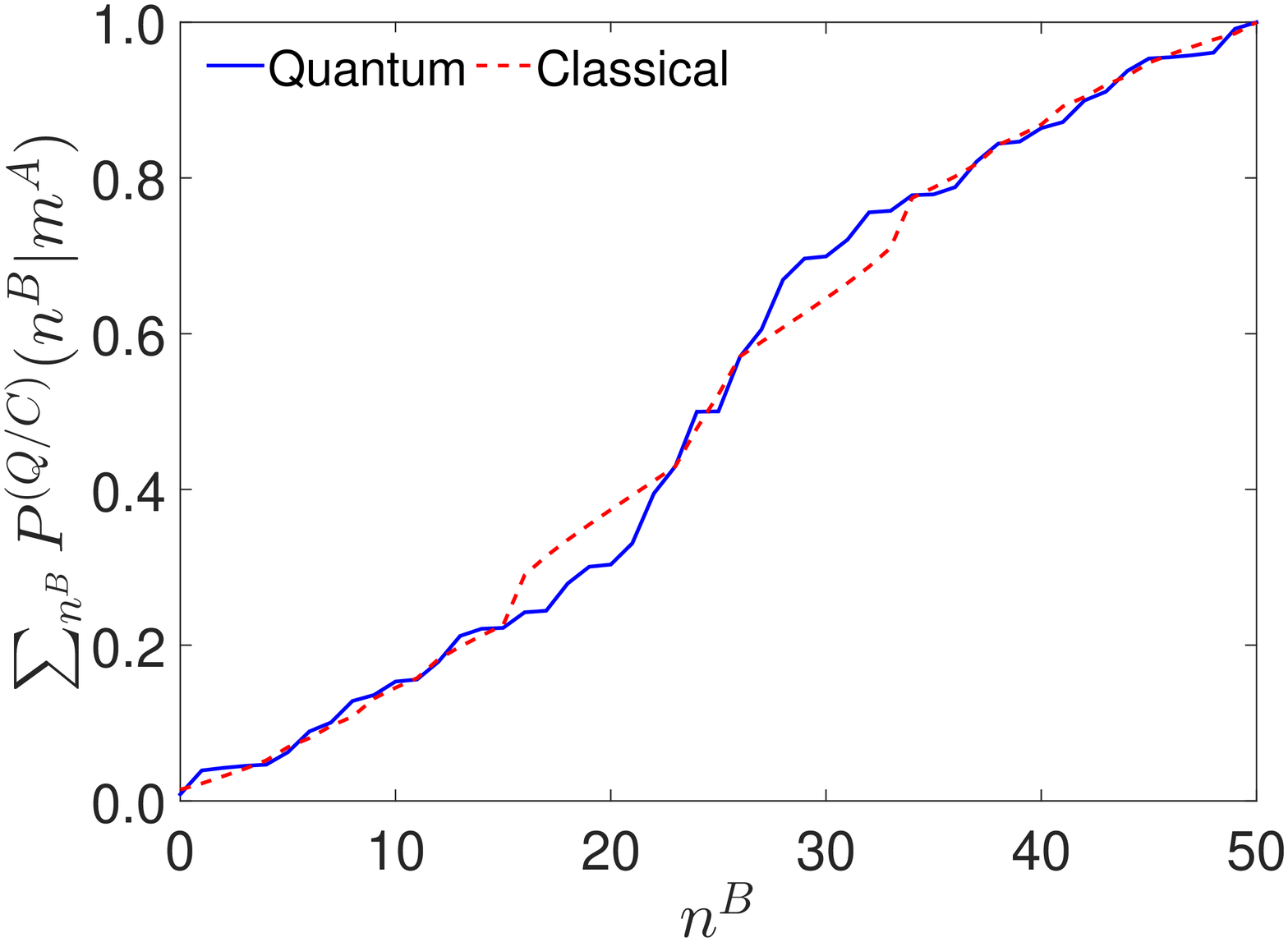}}
  \subfigure[]{\label{fig:subfig:c}
  \includegraphics[width=0.485\textwidth]{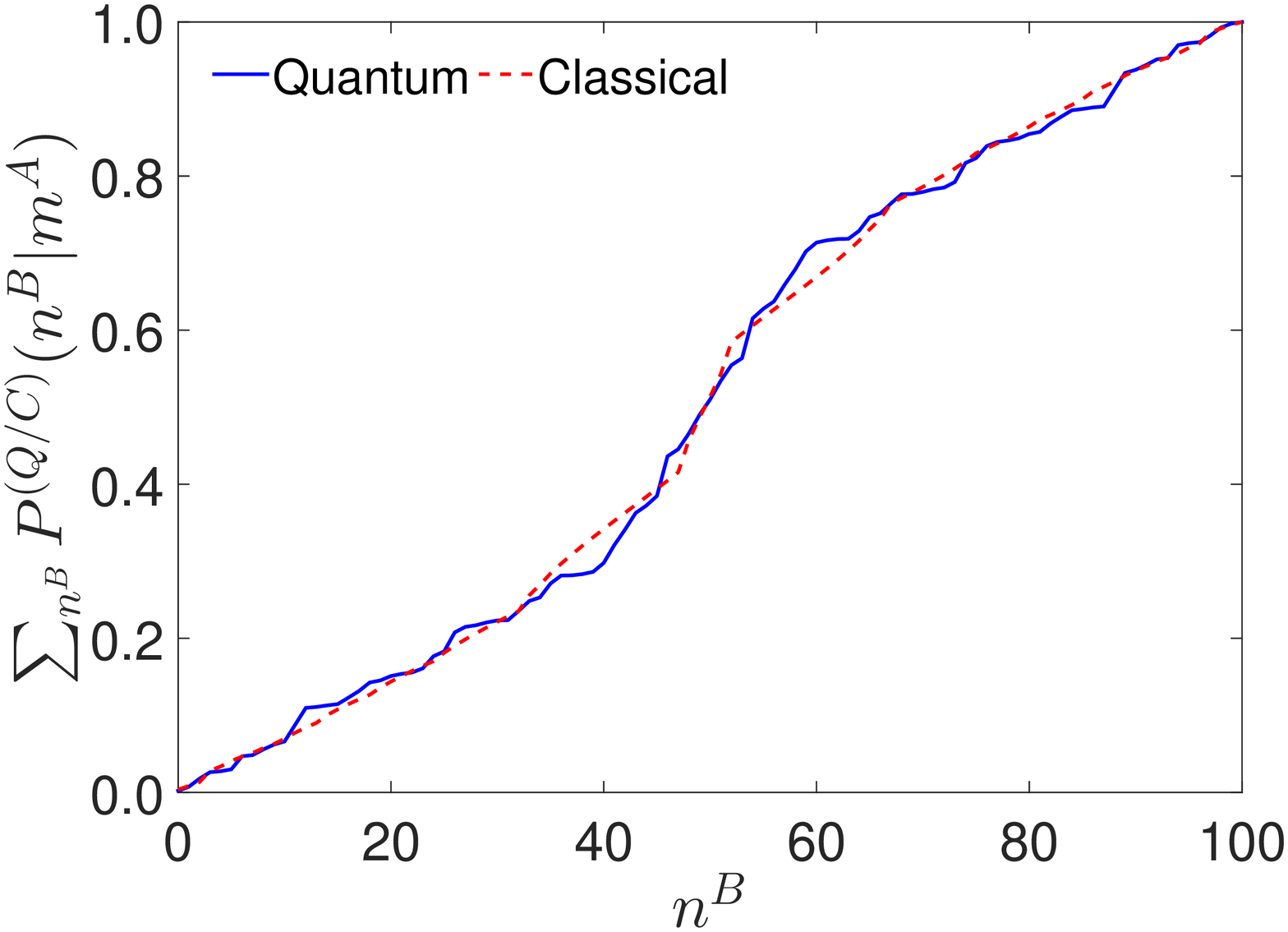}}
  \subfigure[]{\label{fig:subfig:d}
  \includegraphics[width=0.485\textwidth]{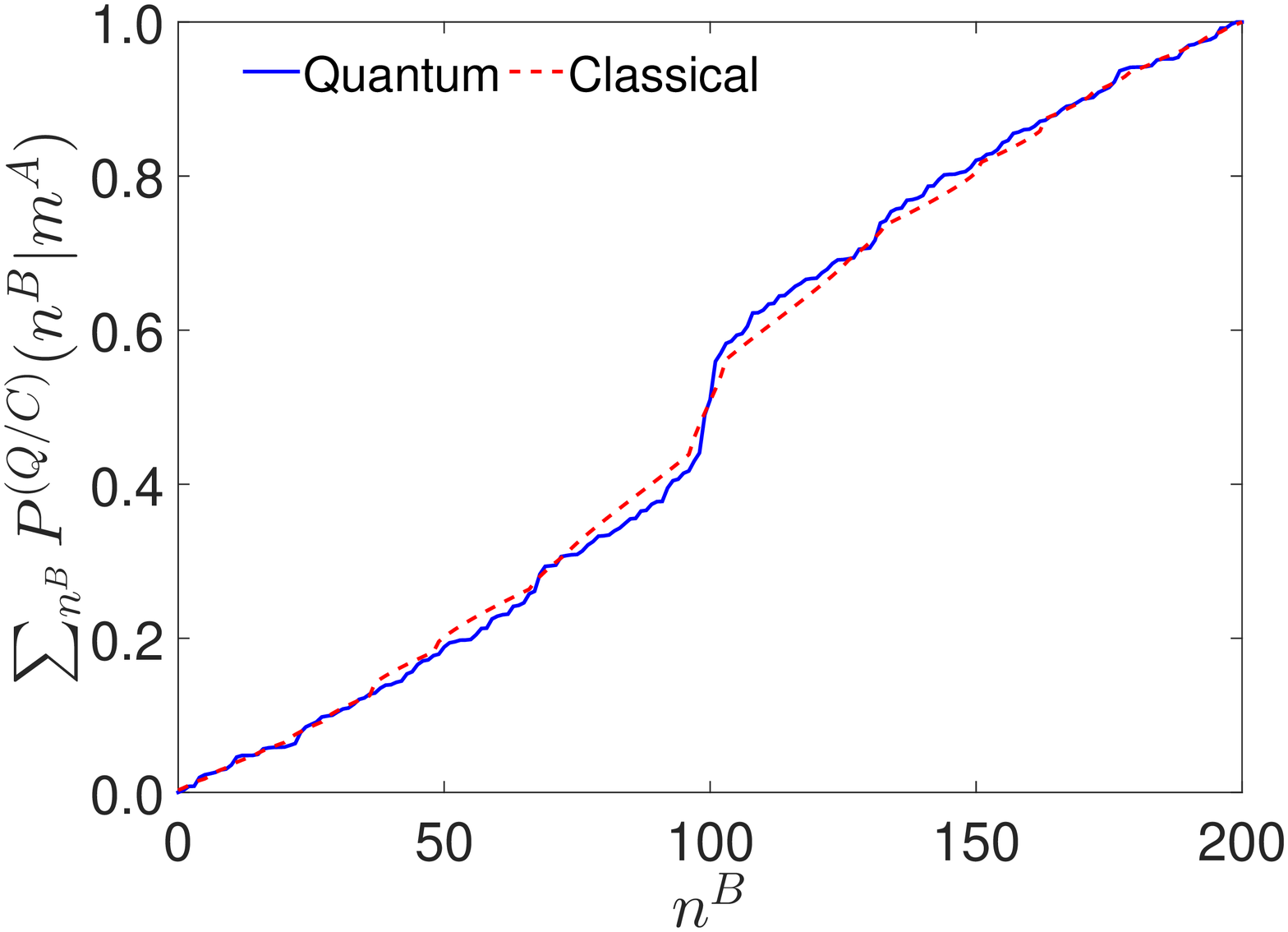}}
  \caption{(Color online) Cumulative quantum and classical transition probabilities for the 1D two-site BH model with different number of particles:
   (a) $N=10$, (b) $N=50$, (c) $N=100$, (d) $N=200$.
   The jagged blue solid curve shows the quantum case, $\sum_{n^B}P^Q(n^B|m^A)$,
   while the smooth dashed red curve shows the classical case, $\sum_{n^B}P^C(n^B|m^A)$. For the quantum case, the initial state is chosen
   to be the ground state of $H(t=0)$ with $|m^A\ra=|\mathbf{n}^A\ra=|N/2,N/2\ra$. The corresponding classical initial state is a collection of microscopic states
   $\Psi^A=\{\psi_1^A,\psi_2^A\}=\{(N/2,\phi_1^A)(N/2,\phi_2^A)\}$, where $\phi_1^A$ and $\phi_2^A$ are the uniformly distributed
   random numbers in the range $[0,2\pi)$.
   Due to the fact that $J(t=\tau)=0$, the final energy eigenstates are Fock states: $|n^B\ra=|n_1^B,N-n_1^B\ra$ with $n_1^B=0,\ldots,N$.}
  \label{acp}
 \end{figure*}

\section{Quantum, semiclassical and classical transition probabilities} \label{QCTP}

 Consider a quantum system, described by a Hamiltonian $\hat{H}(J)$, where $J$ is an externally controlled parameter, usually called
 the work parameter of the system in the field of nonequilibrium statistical mechanics \cite{jarzynski}.
 We study the time evolution of the system when the work parameter $J$ is varied with time from initial value $J(t=0)=A$ to the finial
 value $J(t=\tau)=B$.
 We assume that the the system at $t=0$ is in a thermal equilibrium state at an inverse temperature $\beta$.
 The system is then detached from the heat bath and work is applied when the work parameter $J$ is varied.
 Then following the definition of quantum work \cite{pte}, the work distribution of this nonequilibrium process is given by \cite{pte,chs}
 \be \label{QWD}
   P^Q(W)=\sum_{n^B,m^A}P^Q(n^B|m^A)P^Q(m^A)\delta(W-E_n^B+E_m^A),
 \ee
 where $E_n^B$ and $E_m^A$ are the $n$th and the $m$th eigenvalues of the final and initial Hamiltonian $\hat{H}(t=\tau)$, $\hat{H}(t=0)$, respectively.
 And the corresponding eigenstates are given by $|n^B\rangle$ and $|m^A\rangle$, respectively.
 $P^Q(m^A)$ is the probability of sampling the $m$th eigenstate of $\hat{H}(t=0)$ from the initial thermal equilibrium state when making the initial energy measurement:
 \be
   P^Q(m^A)=\frac{1}{Z_A^Q}\exp\left[-\beta E_m^A\right],
 \ee
 with $Z_A^Q=\sum_m\exp[-\beta E_m^A]$. Given the initial $m$th eigenstate of $\hat{H}(t=0)$, the conditional probability of obtaining the $n$th
 eigenstate of $\hat{H}(t=\tau)$ is given by the quantum transition probability
 \be
   P^Q(n^B|m^A)\equiv|\la n^B|\hat{U}(\tau)|m^A\ra|^2,
 \ee
 with
 \be
    \hat{U}(\tau)=\hat{\mathcal{T}}\exp\left[-\frac{i}{\hbar}\int_0^\tau dt\hat{H}(t)\right],
 \ee
 where $\hat{\mathcal{T}}$ is the time ordering operator.

 For the classical case, we can follow the same lines as we do in the quantum case, except that we are now in the phase space instead of
 the Hilbert space.
 The classical work distribution can be expressed in the following form \cite{chs}:
 \be \label{CWD}
   P^C(W)\approx\sum_{n^B,m^A}P^C(n^B|m^A)P^C(m^A)\delta(W-E_n^B+E_m^A),
 \ee
 where $P^C(n^B|m^A)$ and $P^C(m^A)$ are the classical counterparts of $P^Q(n^B|m^A)$ and $P^Q(m^A)$, respectively.

 With Eqs.~(\ref{QWD}) and (\ref{CWD}),
 the classical and quantum work distributions can be compared directly.
 We begin with comparing the classical and quantum initial probabilities $P^C(m^A)$ and $P^Q(m^A)$.
 Following Ref.~\cite{chs}, we know that the initial distribution for a $d$-dimensional classical system reads
 \be
   P^C(m^A)=\int_{E_m^A}^{E_{m+1}^A}\frac{1}{Z_A^C}\bar{\rho}(E)e^{-\beta E}dE,
 \ee
 where
 \begin{align}
    Z_A^C=\int\frac{d^dp d^dq}{(2\pi\hbar)^d}\exp[-\beta\mathcal{H}_c(p,q)],
 \end{align}
 is the classical partition function and $\bar{\rho}(E)$ is the density of states (DOS) of the classical system.
 For the BH model which we study here, $\bar{\rho}(E)$ has the following expression \cite{englE}:
 \begin{align}
   \bar{\rho}(E)=&\left(\frac{4}{\pi}\right)^L\int d^Lp d^Lq\delta[E-\mathcal{H}_c(\mathbf{p},\mathbf{q})] \notag \\
                 &\times\delta\left(\mathbf{p}^2+\mathbf{q}^2-N-\frac{L}{2}\right),
 \end{align}
 with $\psi_j=q_j+ip_j$ and $\mathbf{q}=(q_1,\ldots,q_L)$, $\mathbf{p}=(p_1,\ldots,p_L)$.

 For the quantum case, $P^Q(m^A)$ has the same form as the classical case except that the partition function is
 given by quantum expression and the DOS now reads
 \be
    \rho(E)=\sum_n\delta(E-E_n),
 \ee
 where $E_n$ are the eigenvalues of the Hamiltonian in Eq.~(\ref{BHH}). According to Gutzwiller \cite{gutzwiller}, in the semiclassical
 limit, i.e., $\hbar\to0$, $\rho(E)$ has the generic form \cite{englE}
 \be
    \rho(E)=\bar{\rho}(E)+\tilde{\rho}(E).
 \ee
 Here the smooth part $\bar{\rho}(E)$ is purely classical, known as the Weyl term, while the oscillatory part $\tilde{\rho}(E)$
 comes from the quantum fluctuations and can be expressed in terms of classical quantities, which are encoded in the classical periodic orbits.
 In our study, the energy scale that we consider is much larger than the periodicity of $\tilde{\rho}(E)$, therefore, we can ignore the oscillatory
 part and approximately take the DOS of the quantum system to be $\bar{\rho}(E)$. Finally, we find that the initial distributions
 of the quantum and classical cases are approximately equal \cite{chs,zhulong}:
 \begin{align}
    P^Q(m^A)&=\int_{E^A_m}^{E^A_{m+1}} dE\frac{1}{Z_A^Q}\rho(E)e^{-\beta E}  \\
            &\approx\int_{E_m^A}^{E_{m+1}^A} dE\frac{1}{Z_A^C}\bar{\rho}(E)e^{-\beta E}=P^C(m^A).
 \end{align}
 Thus, in order to compare the quantum and classical work distributions, the only thing one needs to clarify is the relationship between
 the classical and quantum transition probabilities $P^C(n^B|m^A)$ and $P^Q(n^B|m^A)$.
 In the following, we study these transition probabilities in the 1D BH model explicitly.

 In our study, we change $J$ from $J(t=0)=0$ to $J(t=\tau)=0$.
 Therefore, $A=0$, $B=0$, and both the initial and the final energy eigenstates are given by the Fock states.
 The transition probability between different energy eigenstates is given by the transition probability
 between different Fock states.
 The classical counterpart of the Fock state is a collection of microscopic states $\Psi^A\equiv\{\psi_1^A,\ldots,\psi_L^A\}=\{(n^A_1,\phi^A_1),\ldots,(n_L^A,\phi_L^A)\}$,
 with $n^A_j$'s equal to the number of particles on the $j$th site
 and $\phi^A_j$'s are the independent random numbers which have a uniform distribution in the range $[0,2\pi)$.

 \subsection{Quantum transition probability}

 In order to calculate the quantum transition probability, we expand the wave function evolving under $\hat{H}[J(t)]$ as follows:
 \be \label{swf}
     |\Phi(t)\rangle=\sum_{n_2,\ldots,n_L}^Nc_{n_1,n_2,\ldots,n_L}(t)|n_1,\ldots,n_L\rangle,
 \ee
 where $|n_1,\ldots,n_L\rangle$ are the Fock basis, and the sum is constrained by $N=\sum_{j=1}^L n_j$.
 Therefore, the particle number on the first lattice site is given by $n_1=N-\sum_{j=2}^Ln_j$.
 $c_{n_1,n_2,\ldots,n_L}$'s are expansion coefficients and satisfy the normalization condition
 \be
     \sum_{n_2,\ldots,n_L}^N|c_{n_1,n_2,\ldots,n_L}(t)|^2=1.
 \ee
 Inserting Eq.~(\ref{swf}) into Schr\"{o}dinger equation
 \be
     i\hbar\frac{\partial}{\partial t}|\Phi(t)\rangle=\hat{H}[J(t)]|\Phi(t)\rangle,
 \ee
 after some algebra, we finally get the equations of these coefficients $c_{n_1,n_2,\ldots,n_L}(t)$:
 \begin{align} \label{tec}
     &i\hbar\dot{c}_{n_1,\ldots,n_j,\ldots,n_L}=\frac{U}{2}\sum_{j=1}^L n_j(n_j-1){c}_{n_1,\ldots,n_j,\ldots,n_L}  \notag  \\
           &-J(t)\sum_{j=1}^L\left(c_{n_1,\ldots,n_j-1,n_{j+1}+1,\ldots,n_L}\sqrt{n_j(n_{j+1}+1)}\right. \notag \\
           &\left.+c_{n_1,\ldots,n_j+1,n_{j+1}-1,\ldots,n_L}\sqrt{(n_j+1)n_{j+1}}\right),
 \end{align}
 where the dot denotes the time derivative.
 The quantum transition probability between different Fock states, which we denote by $P^Q(\mathbf{n}^B|\mathbf{n}^A)$
 with $|\mathbf{n}^{A/B}\ra=|n_1^{A/B},\ldots,n_L^{A/B}\ra$, reads
 \be \label{QTPF}
     P^Q(\mathbf{n}^B|\mathbf{n}^A)=|c_{n_1,n_2,\ldots,n_L}(\tau)|^2,
 \ee
 where $c_{n_1,n_2,\ldots,n_L}(\tau)$ solves Eq.~(\ref{tec}) with the initial condition given by $c_{n_1,n_2,\ldots,n_L}(0)$.
 These results will be used in Sec.~\ref{NRS}.

 \subsection{Semiclassical and classical transition probabilities}

 According to Refs.~\cite{engl,englt}, one can write down the semiclassical transition probability between different Fock states of
 the BH model as follows:
 \be
   P^{\mathrm{SC}}(\mathbf{n}^B|\mathbf{n}^A)=|K^{\mathrm{SC}}(\mathbf{n}^{B},\tau;\mathbf{n}^{A},0)|^2,
 \ee
 where $K^{\mathrm{SC}}(\mathbf{n}^{B},\tau;\mathbf{n}^{A},0)$ is the semiclassical propagator, and given by \cite{engl,englt}
 \begin{align}
   K^{\mathrm{SC}}&(\mathbf{n}^{B},\tau;\mathbf{n}^{A},0)   \nonumber  \\
               &=\sum_\gamma\sqrt{\mathrm{det}'\frac{1}{(-2\pi i\hbar)}\frac{\partial^2
               R^\gamma(\mathbf{n}^B,\tau;\mathbf{n}^A,0)}{\partial\mathbf{n}^B\partial\mathbf{n}^A}} \nonumber \\
               &\times\exp\left[\frac{i}{\hbar}R^\gamma(\mathbf{n}^{B},\tau;\mathbf{n}^{A},0)+i\mu^\gamma\frac{\pi}{2}\right].
 \end{align}
 Here, $\gamma$ indexes all classical trajectories satisfying Eq.~(\ref{CTE}) and the boundary conditions
 \bey
   &&|\psi_j(t=0)|^2=n^A_j+\frac{1}{2},  \\
   &&|\psi_j(t=\tau)|^2=n_j^B+\frac{1}{2},
 \eey
 with $j=1,\ldots,L$ and $\arg\psi_1(t=0)=0$
 and $\mu^\gamma$ denotes the Maslov index of the $\gamma$th trajectory, while
 $R^\gamma(\mathbf{n}^B,\tau;\mathbf{n}^A,0)$ is the classical action of the $\gamma$th trajectory
 \be
    R^\gamma(\mathbf{n}^B,\tau;\mathbf{n}^A,0)=
         \mathlarger{\int}_0^\tau\Big[\sum_j\phi_j^\gamma(t)\dot{n}_j^\gamma(t)-\mathcal{H}^\gamma_c(t)/\hbar\Big]dt.
 \ee
 The derivatives of the action $R^\gamma$ with respect to $n_j^A$ and $n_j^B$ are
 \begin{align}
   \frac{\partial R^\gamma(\mathbf{n}^B,\tau;\mathbf{n}^A,0)}{\partial n_j^A}&=-\hbar\phi_j^\gamma(0),  \label{Drf}\\
   \frac{\partial R^\gamma(\mathbf{n}^B,\tau;\mathbf{n}^A,0)}{\partial n_j^B}&=\hbar\phi_j^\gamma(\tau).
 \end{align}
 The prime in the determinant
 \be
    \mathrm{det}'\left(\frac{\partial^2 R^\gamma}{\partial\mathbf{n}^A\partial\mathbf{n}^B}\right)
      \equiv\mathrm{det}\left(\frac{\partial^2 R^\gamma}{\partial n_j^A\partial n_k^B}\right)_{j,k=2,\ldots,L}
 \ee
 indicates that the derivatives skip the first component.
 This is a consequence of the conservation of the total number of particles \cite{engl,englt}.

 Following the same procedure as in Ref.~\cite{chs}, we can further simplify the expression of the transition
 probability by ignoring the interference terms
 between different classical trajectories \cite{englt}
 \be \label{CP}
   P^{\mathrm{SC}}(\mathbf{n}^B|\mathbf{n}^A)\overset{\mathrm{diag}}{\approx}
               \left(\frac{1}{2\pi\hbar}\right)^{L-1}\sum_\gamma\left|\mathrm{det}'\left[\frac{\partial\bm{\phi}(0)}
               {\partial\mathbf{n}^B}\right]\right|,
 \ee
 where $\bm{\phi}(0)$ represents the vector of the initial phases for the $\gamma$th trajectory, and has been obtained in Eq.~(\ref{Drf}).
 For the classical case, the transition probability is given by \cite{englt}
 \begin{align}
   P^C(\mathbf{n}^B|\mathbf{n}^A)=\mathlarger{\int}_0^{2\pi}d^{L-1}\phi^A\prod_{j=2}^L\delta\left[|\psi_j(\mathbf{n}^A,\bm{\phi}^A;\tau)|^2\right. \notag \\
                                  \left.-(n_j^B+1/2)\right].  \label{scpr}
 \end{align}
 Using the property of $\delta$ function, Eq.~(\ref{scpr}) can be rewritten as \cite{engl}
 \be \label{RCp}
   P^C(\mathbf{n}^B|\mathbf{n}^A)=\left(\frac{1}{2\pi\hbar}\right)^{L-1}
        \sum_\gamma\left|\mathrm{det}'\left[\frac{\partial\bm{\phi}(0)}{\partial\mathbf{n}^B}\right]\right|.
 \ee
 By comparing Eqs.~(\ref{CP}) and (\ref{RCp}), we find that the semiclassical transition probability (\ref{CP}) converges to the classical
 transition probability (\ref{RCp}) after taking the diagonal approximation \cite{chs,berry,doron,baranger}.
 Thus, similar to the single-particle system \cite{chs,zhulong}, we have analytically proved that the quantum
 work distribution will converge to the classical work distribution in a quantum many-body system when
 ignoring the interference effect of different classical trajectories.
 In the following we will provide some numerical results of both quantum and classical transition probabilities to demonstrate our central result.

 \begin{figure}
  \includegraphics[width=\columnwidth]{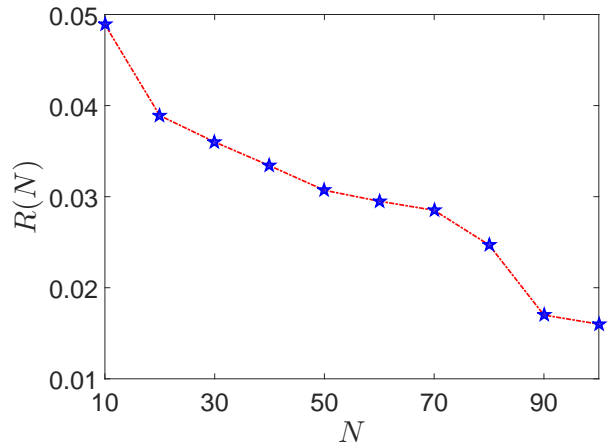}
  \caption{(Color online) RMSE $R(N)$ (blue pentagrams) as a function of the number of particles $N$.
   The other parameters are $U=5/N$, $\tau=10$, $\hbar=1$.}
  \label{rvn}
 \end{figure}

 \section{Numerical results} \label{NRS}

 In this section, we give our numerical results of the 1D two-site and three-site BH models.
 We set $\hbar=1$, $U=5/N$, $\tau=10$ and vary the work parameter $J$ according to the following protocol:
 \be
   J(t)=J_0\left(t-\frac{t^2}{\tau}\right),
 \ee
 with $J_0=5$.
 In our study we also set the particle number $N$ to be an even number.
 Here, we stress that qualitatively similar results can be obtained for any $L$-site BH model with $L\geq2$.

 To calculate the quantum transition probability between different Fock states, we first use a Runge-Kutta
 method to solve the set of coupled ordinary
 differential equations given by Eq.~(\ref{tec}), then use Eq.~(\ref{QTPF}) to obtain the quantum transition probability.
 For the classical case, the shooting method \cite{Whp} has been employed to find all classical trajectories
 from $|\mathbf{n}^A\ra$ to $|\mathbf{n}^B\ra$ at the fixed transit time $\tau$. Then we calculate the classical transition probability
 via Eq.~(\ref{RCp}).

 \subsection{1D Two-site Bose-Hubbard model}

 In this section we study the transition probability in the 1D two-site BH model without periodic boundary condition
 \be \label{twoBH}
   \hat{H}=-J(\hat{a}_1^\dag\hat{a}_2+\hat{a}_2^\dag\hat{a}_1)+
            \frac{U}{2}(\hat{a}_1^\dag\hat{a}_1^\dag\hat{a}_1\hat{a}_1+\hat{a}_2^\dag\hat{a}_2^\dag\hat{a}_2\hat{a}_2).
 \ee
 This is an extensively studied \cite{castin,mck,gce,lena,lena1,kss,ebm,urf,mhs,zibold,rws,rgm,rfv,sra,
 ass,emg,kwm,leggett,gsp,hvs,apt,gkd,jja,vss,fnj,kpp} paradigmatic model and can be realized in various systems,
 for example, particles in a harmonic well \cite{zibold}.
 Under the well-known two-mode approximation, the 1D two-site BH Hamiltonian in Eq.~(\ref{twoBH}) can also be used to describe the dynamics
 of an atomic Bose-Einstein condensate in a double-well potential \cite{gce,rws,rgm}.

 \begin{figure}
  \includegraphics[width=\columnwidth]{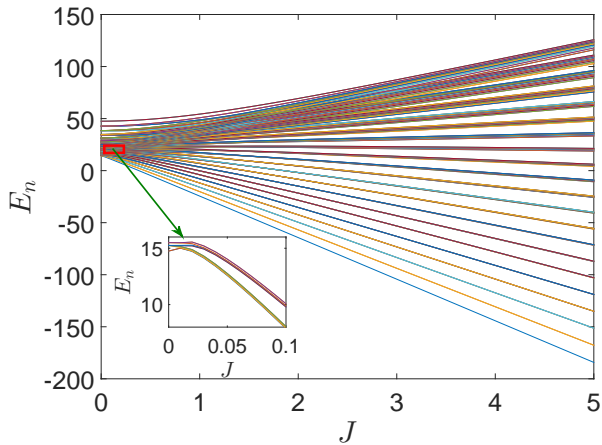}
  \caption{(Color online) Energy spectrum of the 1D three-site BH model (\ref{TBHM}) with the work parameter $J$ for $N=20$.
   Inset: The details of the red rectangle.}
  \label{3energy}
 \end{figure}

\begin{figure*}
  \begin{minipage}[t]{0.5\linewidth}
    \centering
    \includegraphics[width=\columnwidth]{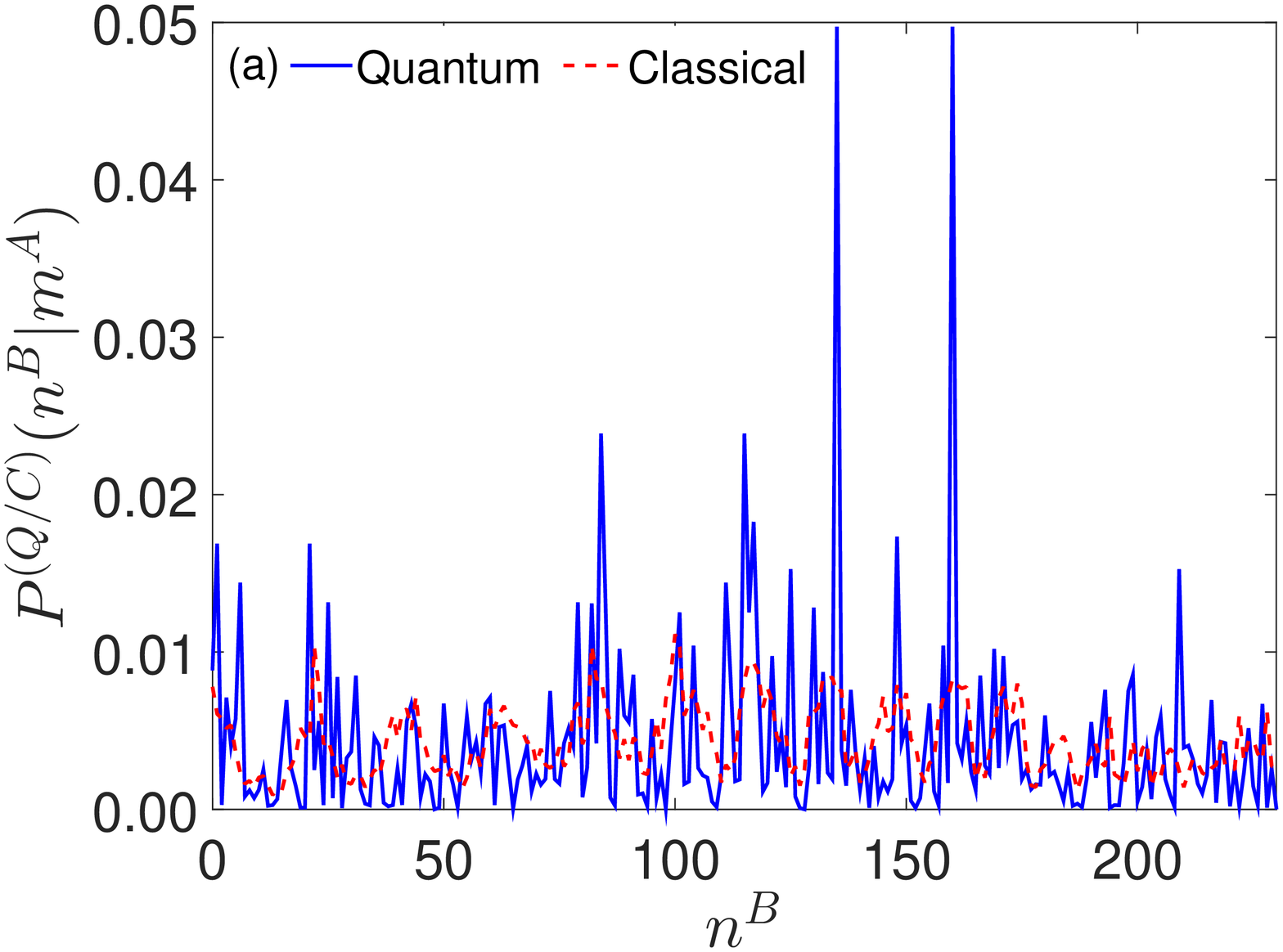}
  \end{minipage}%
  \hfill
  \begin{minipage}[t]{0.5\linewidth}
    \centering
    \includegraphics[width=\columnwidth]{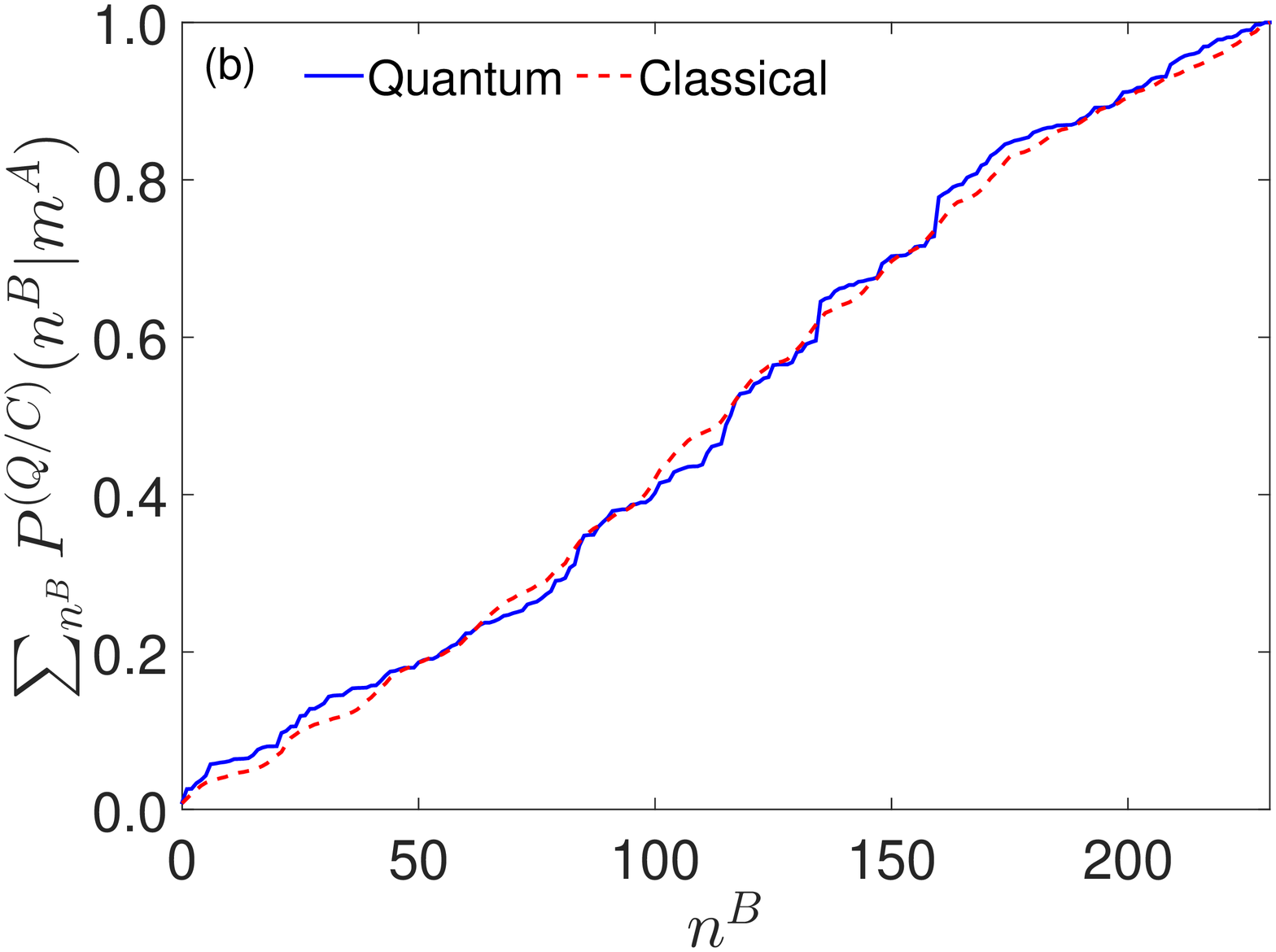}
  \end{minipage}%
  \caption{(Color online) Quantum (solid blue curve) and classical (dashed red curve) transition probabilities
  of the 1D three-site BH model:
  (a) Transition probabilities between different energy eigenstates, (b) Cumulative transition probabilities.
  For the quantum case, the number of bosons is $N=20$ and the initial state is one of the three degenerate eigenstates
  of the $19$th energy level of $H(t=0)$ with $|m^A\ra=|\mathbf{n}^A\ra=|5,5,10\ra$.
  The classical counterpart of $|m^A\ra$ is a collection of microscopic states $\Psi^A=\{\psi_1^A,\psi_2^A,\psi_3^A\}=\{(5,\phi_1^A),(5,\phi_2^A),(10,\phi_3^A)\}$,
  with $\phi_j$'s $(j=1,2,3)$ the uniformly distributed random numbers in the range $[0,2\pi)$.
  Due to the fact that the final value of $J$ is zero, the energy eigenstates of $H(t=\tau)$ are the Fock states: $|n^B\ra=|n_1^B,n_2^B,N-n_1^B-n_2^B\ra$
  with $n_1^B=0,n_2^B=0,\ldots,N; n_1^B=1,n_2^B=0,\ldots,N-1;\ldots;n_1^B=N,n_2^B=0$.}
  \label{prb3}
 \end{figure*}

 We choose the ground state of $\hat{H}(t=0)$ as our initial state. The corresponding Fock state is the
 twin-Fock state, therefore, we have $|\mathbf{n}^A\ra=|N/2,N/2\ra$.
 Its classical counterpart is a collection of microscopic states $\Psi^A=\{\psi_1^A,\psi_2^A\}=\{(N/2,\phi_1^A),(N/2,\phi_2^A)\}$,
 with $\phi_1^A$ and $\phi_2^A$ the uniformly distributed random numbers in the range $[0,2\pi)$.
 Here we should point out that for small $J$ all excited energy levels are doubly
 degenerate and it splits with the increase of $J$ (see Fig.~\ref{energy}).
 However, the quantity that we studied is the transition probability between different energy eigenstates, therefore we do not need to consider
 the effect of the degeneracy.
 Due to the fact that both the initial and the final values of $J$ are equal to zero, the Fock states $|\mathbf{n}^{A/B}\ra$ are also
 the energy eigenstates at the initial and the final moments.
 Hence, the quantum and classical transition probabilities between different energy eigenstates can be expressed as
 the transition probabilities between different Fock states:
 \bey
     &&P^Q(n^B|m^A)=P^Q(\mathbf{n}^B|\mathbf{n}^A), \label{QTP}  \\
     &&P^C(n^B|m^A)=P^C(\mathbf{n}^B|\mathbf{n}^A). \label{CTP}
 \eey
 Here, the relation between the energy eigenstates $|m^{A}\ra$, $|n^B\ra$ and the Fock states $|\mathbf{n}^{A}\ra$, $|\mathbf{n}^B\ra$
 are defined in the captions of Figs.~\ref{qcp}, \ref{acp}, and \ref{prb3}.

 In Fig.~\ref{qcp}, we plot the quantum transition probability for different number of particles
 as a function of the final energy eigenstates $|n^B\ra$ (solid line).
 Comparing with the classical case (dashed line), we find that the quantum probability oscillates rapidly with
 $n^B$. This feature has an origin in the wave nature of the quantum system.
 Obviously, the correspondence between $P^Q(n^B|m^A)$ and $P^C(n^B|m^A)$ is visually evident.

 In order to smooth out the rapid oscillations and to compare these two probabilities in a better way, we plot
 the {\emph{cumulative transition probabilities}}
 $\sum_{n^B}P^Q(n^B|m^A)$ and $\sum_{n^B}P^C(n^B|m^A)$
 in Fig.~\ref{acp} for different number of particles $N$.
 Obviously, the agreements between these two probabilities are not very good for small $N$,
 but the convergence is improved when $N$ increases.
 The deviation observed in small $N$ can be explained as follows: when the number of particles $N$ is small,
 the characteristic actions of the system are not much larger than the effective Planck's constant $\hbar_{\mathrm{eff}}$.
 Therefore, the classical approximations adapted in Sec.~\ref{BHM} [cf. Eqs.~(\ref{trf})-(\ref{CTE})] are expected to be a poor approximation.

 We can also see that the jagged quantum cumulative transition probability oscillates around the
 classical cumulative transition probability. This phenomenon stems from the interference between different classical
 trajectories \cite{chs}.
 The convergence displayed in Fig.~\ref{acp} suggests that there indeed exists a correspondence principle
 between quantum and classical work distributions, despite the nonclassical feature visible in Fig.~\ref{qcp}.

 The convergence of the quantum and classical transition probabilities depends on the number of particles $N$ (see Fig.~\ref{acp}).
 In order to understand the correspondence of work distribution in a better way,
 we use the root-mean-square error (RMSE) \cite{mhd} to quantify the difference between the quantum and classical cumulative probabilities.
 For certain $N$, the RMSE, which we denote by $R(N)$, between these two cumulative probabilities is given by
 \be
   R(N)\equiv\sqrt{\frac{1}{M}\sum_{l=0}^{N}\left[S_l^Q(N)-S^C_l(N)\right]^2},
 \ee
 where $M=N+1$ represents the total number of eigenstates and
 \be
   S_l^{Q/C}(N)=\sum_{n^B=0}^{l}P^{Q/C}(n^B|m^A),
 \ee
 with $l=0,\ldots,N$.

 The RMSE $R(N)$ quantifies the average deviations between two different probability distributions.
 If two probability distributions are identical, we have $R(N)=0$.
 The closer the two cumulative probability distributions $S_l^Q$ and $S^C_l$ are,
 the smaller $R(N)$ is.
 The vanishing of $R(N)$ implies the correspondence principle \cite{zhulong}.
 Hence, the validity of the correspondence principle can be quantitatively characterized by the vanishing of the RMSE.

 RMSE $R(N)$ as a function of particle numbers $N$ is shown in Fig.~\ref{rvn}.
 It is seen that the value of $R(N)$ decreases with the increase of particle numbers $N$.
 In order to satisfy the classical limit ($N\to\infty$), large $N$ is necessary.
 The behavior of $R(N)$ implies that its value will approach zero when the particle numbers go to infinity,
 i.e.,
 \be \label{rnlimit}
   \lim_{N\to\infty}R(N)\to0.
 \ee
 This is in accordance with the well-known correspondence principle that quantum mechanics and
 classical mechanics give the same result in the classical limit.

 \subsection{1D three-site Bose-Hubbard model}

 The 1D two-site BH model is simple and a special case of BH model, in order to study a general case we
 extend our study to the 1D three-site case. The Hamiltonian of the three-site BH reads
 \be \label{TBHM}
   \hat{H}=-J\sum_{j=1}^3\left(\hat{a}_j^\dag\hat{a}_{j+1}+\hat{a}_{j+1}^\dag\hat{a}_j\right)
   +\frac{U}{2}\sum_{j=1}^3n_j(n_j-1),
 \ee
 where the periodic boundary condition (i.e., a ring geometry) $\hat{a}_{L+1}=\hat{a}_1$ has been assumed.
 The three-site system is a non-integrable system and its energy spectrum (Fig.~\ref{3energy}) is less
 regular than that of the two-site system (Fig.~\ref{energy}).
 The dynamics of its classical counterpart is chaotic due to the nonlinear dynamics in a four-dimensional phase space,
 and its behavior is much richer than the two-site setup \cite{smc,ark,api,ehd,rfvp,rfvp2,knca,mhtk,mhtk2}.

 In our study, we choose the initial state to be one of three degenerate eigenstates of the $19$th energy level of the initial Hamiltonian.
 Its corresponding Fock state is $|\mathbf{n}^A\ra=|5,5,10\ra$.
 The classical counterpart of $|\mathbf{n}^A\ra$ is a collection of microscopic states $\Psi^A=\{\psi_1^A,\psi_2^A,\psi_3^A\}=\{(5,\phi_1^A),(5,\phi_2^A),(10,\phi_3^A)\}$,
 where $\phi_1^A$, $\phi_2^A$, and $\phi_3^A$ are the uniformly distributed random numbers in the range $[0,2\pi)$.
 The classical counterpart of the Hamiltonian (\ref{TBHM}) can be found in Sec.~\ref{BHM}.
 And the classical dynamics of the system satisfies three coupled differential equations of $\psi_j$ $(j=1,2,3)$ [cf. Eq.~(\ref{CTE})].

 Figure \ref{prb3}(a) shows the quantum and classical transition probabilities of the three-site BH model with $N=20$.
 It can be seen that
 unlike the 1D two-site case where the behavior of the classical transition probability is regular,
 in the three-site system the classical transition probability is irregular.
 This phenomenon stems from the fact that the dynamics of the 1D three-site BH model is nonintegrable and becomes more and more chaotic as $\lambda$
 increases.
 Surprisingly, for the three-site BH model, the agreement between the quantum and classical cumulative transition probabilities is very
 good even for small $N$ [see Fig.~\ref{prb3}(b)].

 \section{Conclusions and discussions} \label{DSC}

 The quantum-classical correspondence principle for work distribution in a quantum many-body system, i.e.,
 1D BH model, has been studied in this article.
 Since the initial quantum and classical probability distribution functions are approximately equal,
 the correspondence principle between quantum and classical work distributions is equivalent to the
 correspondence between the quantum and classical transition probabilities
 between different energy eigenstates.
 We first analytically demonstrate the convergence of the quantum and the classical transition probabilities by
 utilizing the analytical expression of the semiclassical propagator between Fock states \cite{engl,englt}, and then we
 numerically calculate the quantum and classical transition probabilities in the two-site and three-site 1D BH models.
 We find that the numerical results agree with the analytic result.

 A direct comparison of the quantum and classical transition probabilities shows that
 the quantum transition probability oscillates rapidly along the classical transition
 probabilities due to the interference of different classical trajectories,
 while the classical transition probability is smooth and continuous for the integrable case and irregular for the nonintegrable case.
 Therefore, the classical and quantum probabilities are manifestly different.
 However, for the cumulative probabilities, we have observed good agreement between them.
 Our results also demonstrate that the convergence, which is characterized by the vanishing of the statistical quantity RMSE,
 between cumulative quantum and classical probabilities
 becomes better with the increase of the particle numbers of the system,
 and vanishes as $N\to\infty$.
 This behavior of RMSE implies that in the classical limit the quantum work distribution converge to the classical work distribution.
 Therefore, there indeed exists a quantum-classical correspondence principle of work distributions in a quantum many-body system,
 even though the indistinguishability and interaction make the properties of quantum work elusive.

 Finally, we stress that the quantum-classical correspondence of the BH models studied in this article is a
 dynamic one \cite{chs,zhulong}, namely, for a system governed by a time dependent Hamiltonian, the quantum and classical
 transition probabilities converge to each other in the classical limit.
 Whereas, the usual studies of the quantum-classical correspondence in the BH models \cite{mck,lena,lena1,emg,smc,api,milburn,trimborn}
 are the static case, where the Hamiltonian of the system is time independent.
 Our work, therefore, complements the previous static correspondence principle in the BH model, which has been studied extensively.
 Furthermore, this work also complements the recent progress established in Refs.~\cite{chs} and \cite{zhulong}, and justifies
 the definition of quantum work via two-point energy measurements in a quantum many-body system.

 \acknowledgements

 H.T.Q. gratefully acknowledges support from the National Science Foundation of China
 under Grants No. 11375012 and No. 11534002, and The Recruitment
 Program of Global Youth Experts of China.


\end{document}